\documentclass{jfm}
\usepackage{graphicx}
\usepackage{amsmath}
\usepackage{epstopdf, epsfig}
\usepackage{dcolumn}
\usepackage{bm}
\usepackage[utf8]{inputenc}
\usepackage[T1]{fontenc}
\usepackage{mathptmx}
\usepackage{xcolor}
\usepackage{url}
\usepackage{hyperref}
\usepackage[normalem]{ulem}
\usepackage{subfigure}

\shorttitle{Equation for Aeroacoustics}
\shortauthor{T. K. Sengupta, A. Sengupta and B. Joshi}

\title{Equation for Aeroacoustics in a Quiescent Environment}

\author{Tapan K. Sengupta\aff{1}
  \corresp{\email{tksengupta@iitism.ac.in}},
  Aditi Sengupta\aff{1}
 \and Bhavna Joshi\aff{1}}

\affiliation{\aff{1}Department of Mechanical Engineering, Indian Institute of Technology Dhanbad, Jharkhand-826 004, India.}

\begin{document}

\maketitle

\begin{abstract}
The perturbation equation for aeroacoustics has been derived in a dissipative medium from the linearized compressible Navier-Stokes equation without any assumption, by expressing the same in spectral plane as in {\it Continuum perturbation field in quiescent ambience: Common foundation of flows and acoustics} Sengupta {\it et al.}, {\it Phys. Fluids},{\bf 35}, 056111 (2023).  The governing partial differential equation (PDE) for the free-field propagation of the disturbances in the spectral plane provides the dispersion relation between wavenumber and circular frequency in the dissipative medium, as characterized by a nondimensional diffusion number. Here, the implications of the dispersion relation of the perturbation field in the quiescent medium are probed for different orders of magnitude of the  generalized kinematic viscosity, across large ranges of the wavenumber and the circular frequency. The adopted global spectral analysis helps not only classify the PDE into parabolic and hyperbolic types, but also explain the existence of a critical wavenumber depending on space-time scales.  
\end{abstract}

\section{Introduction}
There is a long precedence of studies on wave propagation in various branches of physics, and yet there is no clear definition of waves, as noted by \cite{Whitham74}.  The canonical wave equation was first described by \cite{DAlembert1750} as, 

\begin{equation} \label{Eq:utt}
    u_{tt} = c^2 u_{xx}
\end{equation}

 \noindent for the one-dimensional transverse vibration of string in tension. The analytical solution of equation \eqref{Eq:utt}, subject to initial conditions is available in textbooks (see, e.g., \cite{Sengupta2013} and \cite{SenguptaBhum20}), which is characterized by the non-dissipative and non-dispersive nature (i.e. frequency-wavenumber independence) of the solution that is used as a benchmark for developing numerical methods in different branches of engineering and applied physics. For electromagnetic field, \cite{maxwell54, maxwell1865} obtained the wave equations for the electric field $E$, and the magnetic field $B$, with $c$ as the speed of light in a medium of permeability $\mu_p$ and permittivity $\epsilon_p$ that fix $c = 1/\sqrt{\mu_p \epsilon_p}$. An electromagnetic wave is transverse in nature, with $E$ and $B$ oscillating perpendicular to the direction of wave propagation. 

Other notable wave phenomena governed by equation \eqref{Eq:utt} are given in \cite{Mullothetal2015}. The canonical wave equation in acoustics for perturbation pressure is given by \cite{Feynman65, Feynman69}, based on Euler equation. Similarly, elastic wave propagation in solid mechanics in \cite{Whitham74} relates applied strain and stress, with the longitudinal displacement $u$, given by equation \eqref{Eq:utt} and $c^2 = E_0/\rho$, where $E_0$ is the Young's modulus, and $\rho$ is the density of the medium. 

The interest in information propagation as sound and flow perturbations arose in a bid to develop an unified description for the disturbance propagation in a dissipative medium, as initiated in \cite{Senguptaetal2023}. The fundamental governing equation for perturbation pressure in a quiescent medium is given by,

\begin{equation} \label{Eq:12}
\frac{\partial^2 p'}{\partial t^2}  = c^2\nabla^2 p' + \nu_l \frac{\partial }{\partial t} \nabla^2  p'
\end{equation}

\noindent where the generalized kinematic viscosity is defined as, $\nu_l = \frac{\lambda + 2 \mu}{\bar \rho}$, which includes the effects of the losses during the propagation of the signal either as transverse or as longitudinal waves and/ or diffusive disturbances. The derivation of equation~\eqref{Eq:12} does not require the Stokes' hypothesis \cite{stokes1851}; with the first and second coefficient of viscosity contribute to the bulk viscosity for the signal attenuation, specifically for the propagation of acoustic signal as longitudinal wave via compression and dilatation. Equation~\eqref{Eq:12} incorporates the first and second coefficients of viscosity via the bulk viscosity defined by $\mu_b = \lambda + \frac{2}{3} \mu$. As reported in \cite{Senguptaetal2023}, the effects of bulk viscosity for the propagation of the compression and dilatation waves are included for the onset of the Rayleigh–Taylor instability in \cite{Acoustic_POF13, Acoustic_POF14} showing improved solution of compressible Navier-Stokes equation without invoking the Stokes' hypothesis. Readers interested for a comprehensive review of research on the effects of $\mu_b$ are referred to in \cite{Acoustic_POF15} and many other references contained therein. From classical thermodynamic point of view, the Stokes’ hypothesis equates the thermodynamic pressure to the mechanical pressure for fluid flow. Furthermore for gases, violation of Stokes' hypothesis is also related to relaxation processes of vibrational modes of the polyatomic medium. Air containing {\it diatomic nitrogen and oxygen molecules experiences inelastic transfer of energy due to collisions between these molecules, changing the translational energy resulting in changes of the normal stresses, \cite{Senguptaetal2023}}. Numerical estimates of $\mu_b$ for ideal /noble gases and liquids have been reported in \cite{Acoustic_POF16, Acoustic_POF17}. Unlike in the Stokes' hypothesis, $\lambda$ is independent of $\mu$, and that can be orders of magnitude higher in value. Needless to point out that $\lambda$ and $\mu$ are both dissipative, and hence must have positive sign, while Stokes' hypothesis violates this observation. Furthermore, \cite{Acoustic_POF16} has shown that many common fluids and even diatomic gases, display $\mu_b$ as thousand times larger than $\mu$.

In deriving equation~\eqref{Eq:12}, the following polytropic relation between pressure and density perturbations has been presumed,
 
\begin{equation} \label{Eq:10}
	\frac{\partial \rho'}{\partial t} = \frac{1}{c^2}\frac{\partial p'}{\partial t}
\end{equation}

However, for a more generic polytropic processes, one relates the perturbation pressure with the density perturbation as, $p' = K_1 (\rho')^{n}$. Thus, one can rewrite the above equation by, 

\begin{equation} \label{Eq:10a}
	\frac{\partial p'}{\partial t} = \frac{n}{\gamma} c^2 \frac{\partial \rho'}{\partial t}
\end{equation}

For acoustic signal propagation in a perfect gas, the polytropic index, $n$, can be considered to lie between one (for isothermal process, as assumed in the beginning erroneously for sound wave propagation) and $\gamma$ (which is the ratio of specific heats for constant pressure and constant volume processes) for the propagation of acoustic wave as a reversible adiabatic process by neglecting losses and heat transfer. The latter is also associated with the isentropic process assumed for the classical wave equation~\eqref{Eq:utt}. However, the propagation of sound in a dissipative medium, will entail small heat transfer due to losses, and the polytropic index ($n$) will be different from $\gamma$ by a small amount. The small departure from isentropic condition stems from the observation for perfect gas, for which the change of entropy across a normal shock wave is known to vary as the cube of the pressure jump across it. In contrast, during the propagation of acoustic signal via compression and dilatation, the associated pressure jump will be negligibly smaller. Thus in view of reporting the linearized analysis result, one can ignore the difference between $n$ and $\gamma$ in equation~\eqref{Eq:10a}. Moreover, the dispersive nature of propagating sound waves raises concerns about assuming a single value for the speed of sound.

The propagation of noise in air requires procedures in calculating sound absorbed in standard meteorological condition, such as given by standards issued by SAE in 1975 and ANSI in 1978, whose details are available in \cite{Zuckerwar_Meredith1984}. The improved standards were obtained additionally using the concepts of molecular absorption of sound by nitrogen at lower frequencies and vibrational mode of energy exchange between water vapor and oxygen molecules prevalent at higher frequencies.

Studies conducted in the free-field lead to effects of nonstationarity, inhomogeneity and spreading. A comprehensive set of low frequency laboratory measurements needed to identify the relaxation frequency of nitrogen in air was met by \cite{Zuckerwar_Meredith1984} in the experiment performed in a resonant tube. This also provided independent sound absorption measurements overlapping those in dry air, and closing the gap between oxygen relaxation theory and experiment. Thus, one can understand the theoretical properties by studying planar propagation of the perturbation field. 

For the one-dimensional planar propagation of the perturbation field, equation~\eqref{Eq:12} simplifies to (as given in \cite{Blackstock2000} and derived in \cite{Senguptaetal2023}), 

\begin{equation} \label{Eq:13}
\frac{\partial^2 p'}{\partial t^2}  - c^2\frac{\partial^2 p'}{\partial x^2} - \nu_l \frac{\partial^3  p' }{\partial t\partial x^2}  = 0
\end{equation}

The hydrodynamic and acoustic components of the pressure field are noted over a wide magnitude scales. Thus, it is a challenge to solve  simultaneously the flow and acoustic problems. The major contribution in \cite{Senguptaetal2023} is the use of global spectral analysis (GSA) given in \cite{Sengupta2013, Acoustic_POF18, Sagautetal2023} to classify the wavy and non-wavy nature of the solutions of the space-time dependent PDEs, and is based on the necessary condition of the vanishing of the imaginary part of the amplification factors of the latter for all the physical modes. This classification method for PDEs is different from the classical approach available in textbooks, as in \cite{WFAmes, Sengupta2003}. For the ease of understanding the same, a brief description is provided in the following section. 

The paper is formatted in the following manner. In the next section, the analysis of space-time dependent PDEs is described. In section 3, ramifications of this classification are provided for the propagation of acoustic signal in quiescent free-field over an extended ranges of wavenumbers and circular frequencies. Also in this section, the dispersive nature of acoustic signal propagation is further discussed to highlight the difference between the classical wave equation with the space-time dependent perturbation field equation developed and presented here. Specifically, the transformation of the wave equation to diffusion equation across a critical wavenumber is presented in both the sections 3 and 4. The paper closes with a summary and conclusion in section 5. 


\section{Global Spectral Analysis of space-time dependent PDEs}

This is demonstrated with the help of the space-time dependent acoustic signal propagation equation \eqref{Eq:13}. This equation arises from the linearized response for free-field propagation in flows and/ or in acoustics. This is presented in the spectral plane by representing the fluctuating pressure by, 

\begin{equation} \label{Eq:14}
p'(x,t) = \int \int \hat p (k,\omega)e^{i(kx-\omega t)} dk d\omega 
\end{equation}
The governing equation~\eqref{Eq:13} is represented in the spectral plane by using the above to get the dispersion relation as, 
\begin{equation} \label{Eq:15}
\omega ^2  + i\nu_l k^2 \omega - c^2 k^2 = 0 
\end{equation}

However, in the framework of GSA, one rewrites the perturbation pressure via the hybrid representation as, 

\begin{equation} \label{Eq:14GSA}
p'(x,t) = \int_{Br} \hat p (k,t)e^{ikx} dk  
\end{equation}

The integral on the right hand side is performed along the Bromwich contour, $Br$, defined in its strip of convergence as described in \cite{VanderPol_Bremmer, IFTT}. For the GSA of the governing equation, a length-scale ($L_s$) and a time-scale ($\tau_s$) are introduced, so that one can define the physical amplification factor with respect to $\tau_s$ and the wavenumber ($k$) as,

\begin{equation} \label{Eq:Amp_Fac1}
G_{1,2} (k,\tau_s) = \frac{\hat p(k,t+\tau_s)}{\hat p(k,t)}
\end{equation}

To present the results in non-dimensional form, nondimensional wavenumber is introduced as $kL_s$ and a non-dimensional time is introduced by, $N_\tau = c\tau_s/ L_s$. If $L_s$ is the smallest resolved length scale, then $kL_s$ will span from zero to the Nyquist limit of $\pi$ as explained in \cite{Sengupta2013}.  

Using GSA, one can define the physical amplification factor as in \cite{Acoustic_POF18} by,

\begin{equation} \label{Eq:Amp_Fac2}
G_{1,2} (k,\tau_s) = e^{-i\omega_{1,2} \tau_s}
\end{equation}
\noindent where $\omega_{1,2}$ are obtained as the roots of the dispersion relation in equation~\eqref{Eq:15}, given by,  

\begin{equation}\label{Eq:16}
\omega_{1,2} = \frac{-i \nu_l k^2}{2} \pm kcf
\end{equation}

\noindent where, $f = \sqrt{1-\left(\frac{\nu_l k}{2c} \right)^2}$ is the factor that defines the deviation of the dispersion relation from its non-dissipative, isentropic counterpart of the classical wave equation. In GSA, the wavenumber $k$ is the independent variable, and the dispersion relation in equation~\eqref{Eq:15}, provides the dependence of the circular frequency on $k$.
The complex exponents of the amplification factors indicate corresponding phase shifts given by, 
\begin{equation}\label{Eq:19}
\beta_{1,2} = \pm kcf~\tau_s
\end{equation}
The positive value of $k$-dependent $f$ shows the dispersive nature of the dissipative medium, as opposed to the non-dispersive nature of the classical wave equation. The phase speed and the phase shift are related by the nondimensional phase speeds of the acoustic equation as, 
\begin{equation} \label{Eq:21}
    \frac{c_{ph 1,2}}{c} = \frac{\beta_{1,2}}{kc~\tau_s} = \pm f 
\end{equation}

The corresponding group velocity components ($v_{g 1,2}$) of the acoustic equation are obtained as, 

\begin{equation} \label{Eq:22}
    v_{g 1,2} = \frac{d \omega_{1,2}}{dk} = \pm cf \mp \frac{\left(k \nu_l \right)^2}{4fc} - i \nu_l k
\end{equation}

An important aspect of this analysis is explained next with the help of the real and imaginary parts of the first physical amplification factor written as,

\begin{equation} \label{Eq:G1}
(G_{1})_{real} = e^{-\frac{k^2 \nu_l}{2}} \cos (kfc\tau_s), \;\; \;\; (G_{1})_{imag} = -e^{-\frac{k^2 \nu_l}{2}} \sin (kfc\tau_s)
\end{equation}

These two parts indicate a phase shift over $\tau_s$ to be given by $\beta_1$, as noted above.  
This fixes a phase speed and group velocity over this time interval, as given above.  

For a physical system that does not admit {\it anti-diffusion}, it has been shown in \cite{Acoustic_POF18} that the imaginary part of $v_{g 1,2}$ must be zero. However, for a non-dissipative system: $\nu_l = 0$, and then the governing equation becomes non-dispersive, with group velocity equal to the phase speed, as is the case for the classical wave equation, with both of these equal to $c$. Presence of the dissipative term makes the dynamical system diffusion-dominated, and one can define a diffusion number given by, 
$D_n = \nu_l \tau_s / L_s^2$. 

A typical estimate of non-zero bulk viscosity is given in \cite{Senguptaetal2023}, where the general kinematic viscosity has been obtained by the regression analysis of the experimental data in \cite{Acoustic_POF21}. This resulted in $\mu_b = 7.383 \times 10^{-4} + 3.381 \times 10^{-4} T^*$, where $T^*$ is in Kelvin. Considering air at the temperature of $20^0$C and the corresponding density, dynamic viscosity, one obtains the value of $\lambda / \mu = 9520$. For the analysis presented in a limited region of the $(N_\tau, kL_s)$-plane in \cite{Senguptaetal2023}, the properties have been demonstrated for $D_n= 0.14$. In contrast, a detailed analysis is presented here for three values of $D_n = 1.4$, 0.14 and 0.07 in the same region of the $(N_\tau, kL_s)$-plane that brings out important insights on the length- and time-scales' relationship, which shows change of characteristics of the governing PDE from hyperbolic to parabolic type, depending on the length-scale for a fixed time-scale, as explained later for this sub-critical range of wavenumber.

 The ramifications of the GSA in classifying space-time dependent PDEs are described in the next section.

\section{Characterization of space-time dependent PDEs by GSA}
The characterization of the PDE given by equation~\eqref{Eq:12} is explained analytically, by inspection of the amplification factors for different length and time scales. These properties are displayed in the $(N_\tau, kL_s)$-plane in the following for the chosen diffusion number $D_n$, as it was reported in \cite{Senguptaetal2023} for the single case of $D_n =0.14$ only. This way of representing the property for fixed $D_n$ is advantageous, but the results need careful interpretation, as $N_\tau$ and $D_n$ are both functions of $\tau_s$. In the following, properties for three different values of $D_n$'s are presented here. One notices the fact that the dispersion relation given by equation~\eqref{Eq:15} is quadratic, i.e. the dispersion relation for planar wave propagation has two modes.

\begin{figure*}
\centering
\includegraphics[width=0.9\textwidth]{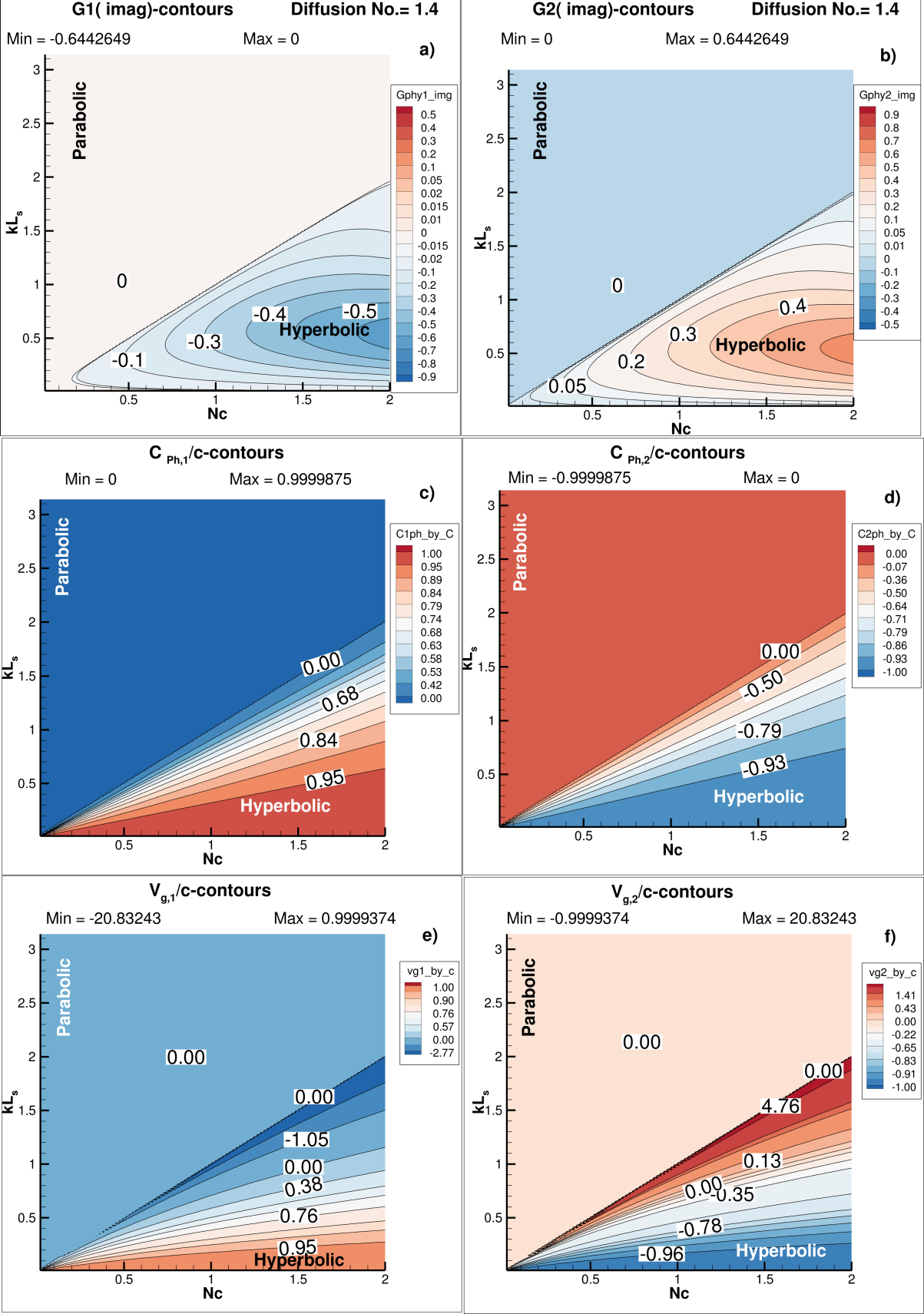}
\caption{The top, middle and bottom frames show for both the modes: imaginary part of the amplification factor, the phase speed, and the group velocity for the diffusion number of $D_n =1.4$.}
\label{Fig1p4}
\end{figure*}

In figure \ref{Fig1p4}, the properties are shown for the case of a large diffusion number, $D_n = 1.4$, for both the modes. In the top frames, the imaginary part of $G_{1,2}$ are shown for this $D_n$, as a function of the nondimensional wave number ($kL_s$), and nondimensional time scale, 
$N_\tau = c \tau_s/ L_s$ combinations. Absence of the imaginary part of $G_{1,2}$ indicates cases when the amplification factor is strictly diffusive. Such a condition is representative of a parabolic PDE, whose amplification factor is strictly real, as shown using GSA in \cite{Sengupta2013}. When the amplification factor is complex in the remaining part of ($N\tau,kL_s$)-plane, the equation \eqref{Eq:12} represents an attenuated wave, which is typical of hyperbolic PDE. From the definition of $D_n$, $N_\tau$ and $k_c = 2c/ \nu_l$, it is readily apparent that $k_c L_s = 2N_\tau /D_n$. Thus, the linear relationship between $k_c L_s$ with $N_\tau$ is readily apparent, with the former given as the boundary between the parabolic and hyperbolic PDE regions. This demarcating slanted straight line is clearly noted for both the modes to be identical, with the slope of the straight line inversely proportional to $D_n$. 

In the middle two frames, the two components of the physical phase speed given in equation~\eqref{Eq:21} are plotted in the same ($N_\tau, kL_s$)-plane for $D_n= 1.4$, and the region adjacent to the $y$-axis has the contour value of $c_{1ph}$ equal to zero, that again implies the parabolic nature of the governing PDE, noted for both the modes. In the hyperbolic part of the domains in ($ N_\tau, kL_s$)-plane, with the red region indicating a right running wave, and the blue contours indicate left running wave. 

In the bottom two frames, the two components of the physical group velocity given in equation~\eqref{Eq:22} are plotted in the same ($N_{\tau}, kL_s$)-plane for $D_n = 1.4$, and the left portion above the $k_c$-line has the group velocity equal to zero, once again implying the parabolic nature of the governing acoustic equation, noted for both the energy carrying modes. In the hyperbolic part of the region in ($N_{\tau}, kL_s$)-plane, the red region indicates right running wave and the blue contours indicate left running wave. The boundary between the parabolic and hyperbolic PDEs are defined from equation~\eqref{Eq:16} for which $f=0$, rendering $\omega_{1,2}$ as strictly imaginary.

\begin{figure*}
\centering
\includegraphics[width=0.9\textwidth]{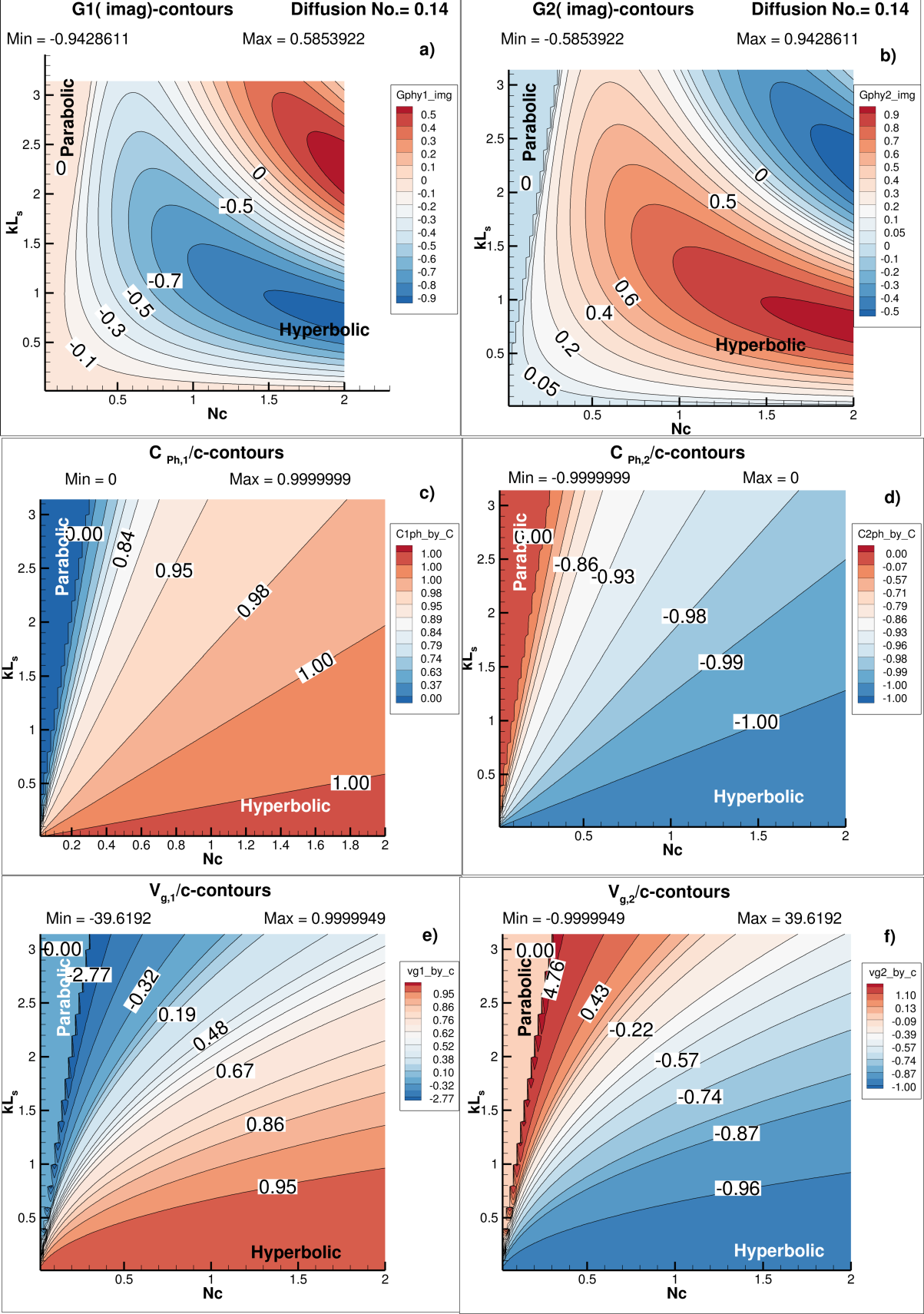}
\caption{The top, middle and bottom frames show for both the modes: imaginary part of the amplification factor, the phase speed, and the group velocity, for the diffusion number of $D_n =0.14$.}
\label{Figp14}
\end{figure*}

In figure \ref{Figp14}, the results are shown for the case of $D_n = 0.14$, to compare with the results in figure~\ref{Fig1p4}. In the top two frames, the imaginary part of $G_{1,2}$ are shown for this $D_n$, in the ($N_{\tau}, kL_s$)-plane for identical ranges. As before, absence of the imaginary part helps identify the values of $kL_s$ and $N_{\tau}$ for which the amplification factors are strictly diffusive. From the relation among $D_n$, $N_{\tau}$ and $k_cL_s$, it is already noted that the demarcating straight line for $k_c L_s$ is sloped more towards the ordinate-axis, for the lower value of $D_n$ in figure~\ref{Figp14}. These non-dimensional sloping straight lines for the non-dimensional critical cut-off wavenumber are noted for both the modes in the contour plots for the phase speed and the group velocity. 

\begin{figure*}
\centering
\includegraphics[width=0.9\textwidth]{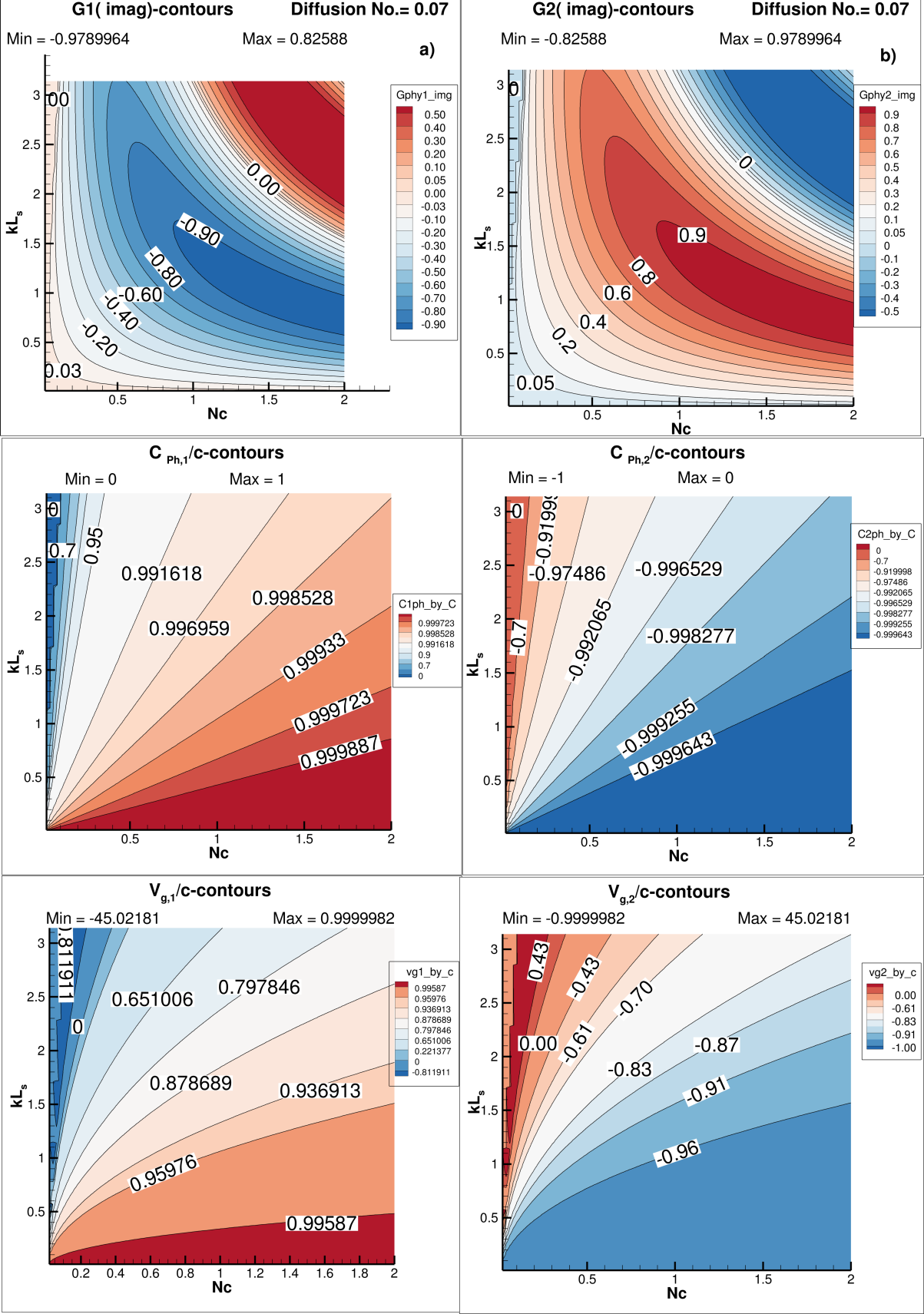}
\caption{The top, middle and bottom frames show for both the modes: imaginary part of the amplification factor, the phase speed, and the group velocity, for the diffusion number of $D_n =0.07$.}
\label{Figp07}
\end{figure*}

In figure \ref{Figp07}, the properties of propagating disturbances are shown for the case of $D_n =0.07$. Once again, the contour plots of the imaginary part of the amplification factors are shown in the top two frames for both the modes in the $(N_\tau, kL_c)$-plane spanning the same ranges of the abscissa and the ordinate. Due to a further lowered value of $D_n$, the critical wavenumber given by the $k_c L_s$-line aligns more closely to the $y$-axis, as noted in all the six frames of this figure. 

For the three values of $D_n$, it is noted that for the hyperbolic solution the first mode has non-dimensional phase speed ranging between 0 and +1 (right running wave), while the second mode shows this non-dimensional phase speed spanning between -1 to 0 (left running wave). The exact value of zero is noted in the range where the governing PDE is parabolic for both the modes. For the group velocity components, the modes show the wave-packet propagation speed to be zero in the parabolic region of the $(N_\tau, kL_s)$-plane, while the hyperbolic part of the region shows non-zero values spanning between negative and positive values. The first mode displays a large negative value for the group velocity, while the positive value is restricted to +1. The magnitude of the negative group velocity increases with decreasing value of $D_n$. similarly, the second mode displays a large positive value, while the negative value is restricted to -1.

\subsection{Alternate description of acoustic signal propagation in a dissipative medium}

So far, we have noted the role of dissipative medium in propagating perturbation to be determined by the length- and time-scales, in showing the linearized governing PDE to be either parabolic or hyperbolic in nature. This is determined by the dispersion relation. The nature of acoustic signal propagation is further discussed to highlight the difference between the classical wave equation with the space-time dependent perturbation field equation developed here. 

\begin{figure*}
\centering
\includegraphics[width=0.9\textwidth]{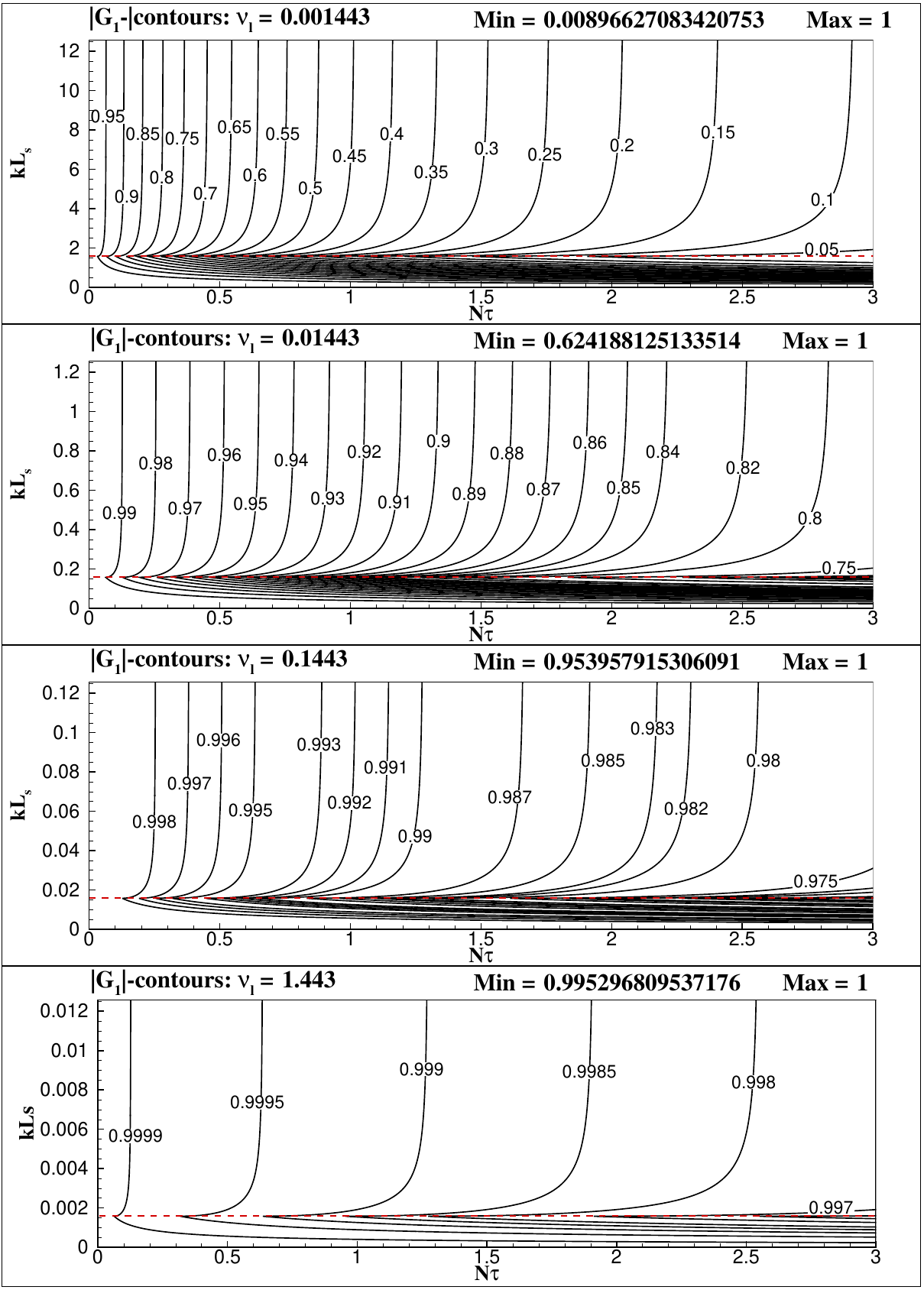}
\caption{The frames show the magnitude of the first mode of the amplification factor for the indicated generalized kinematic viscosity cases in the $(N_\tau, kL_s)$-plane, as given by equation \eqref{Eq:G12_dim}. The range of nondimensional time scale is arbitrary. The boundary between the regions depicting the diffusion and wave solutions given by the cut-off wavenumber ($k_c$) is shown by the dotted line (red). The range of nondimensional wavenumber ($kL_s$) is fixed with $L_s = 3.30485 \times 10^{-4}$m fixed for the case of $D_n =0.14$.}
\label{modG1}
\end{figure*}

In solving equation \eqref{Eq:13} by presenting the properties in terms of the amplification factor (equations~\eqref{Eq:Amp_Fac1}, \eqref{Eq:Amp_Fac2}), the modal phase speeds (equation~\eqref{Eq:21}), and the modal group velocities (equation~\eqref{Eq:22} in figures \ref{Fig1p4}, \ref{Figp14} and \ref{Figp07}, we have used non-dimensional length-scale ($L_s$) and the time-scale  ($\tau_s$), with constant diffusion number ($D_n$) as a parameter. This helped in identifying a critical wavenumber ($k_c$) line in the non-dimensional $(N_\tau, kL_s)$-plane with its equation given by, $k_c L_s = 2N_\tau /D_n$. While each figure plotted for $D_n = constant$ contains wealth of data in the hyperbolic region of the domain, lesser information can be gleaned in the parabolic region. Presence of such a transition from the wave equation to a diffusion equation is similar to the conjecture attributed to Kolmogorov length scale for very high Reynolds number convection dominated flow in turbulent regime, as noted in \cite{Acoustic_POF25}. It has been derived from the first principles in \cite{Senguptaetal2023} and here. However, one also notes the fact that for the constant-$D_n$ plots, each and every point represents a fluid with different generalized viscosity ($\nu_l$). To circumvent these difficulties, it is preferable to represent the properties in the same non-dimensional plane, but for $\nu_l = constant$ as the parameter. This immediately fixes the cut-off wavenumber given by, $k_c = 2c/\nu_l$ as a horizontal straight line. 

\begin{figure*}
\centering
\includegraphics[width=0.9\textwidth]{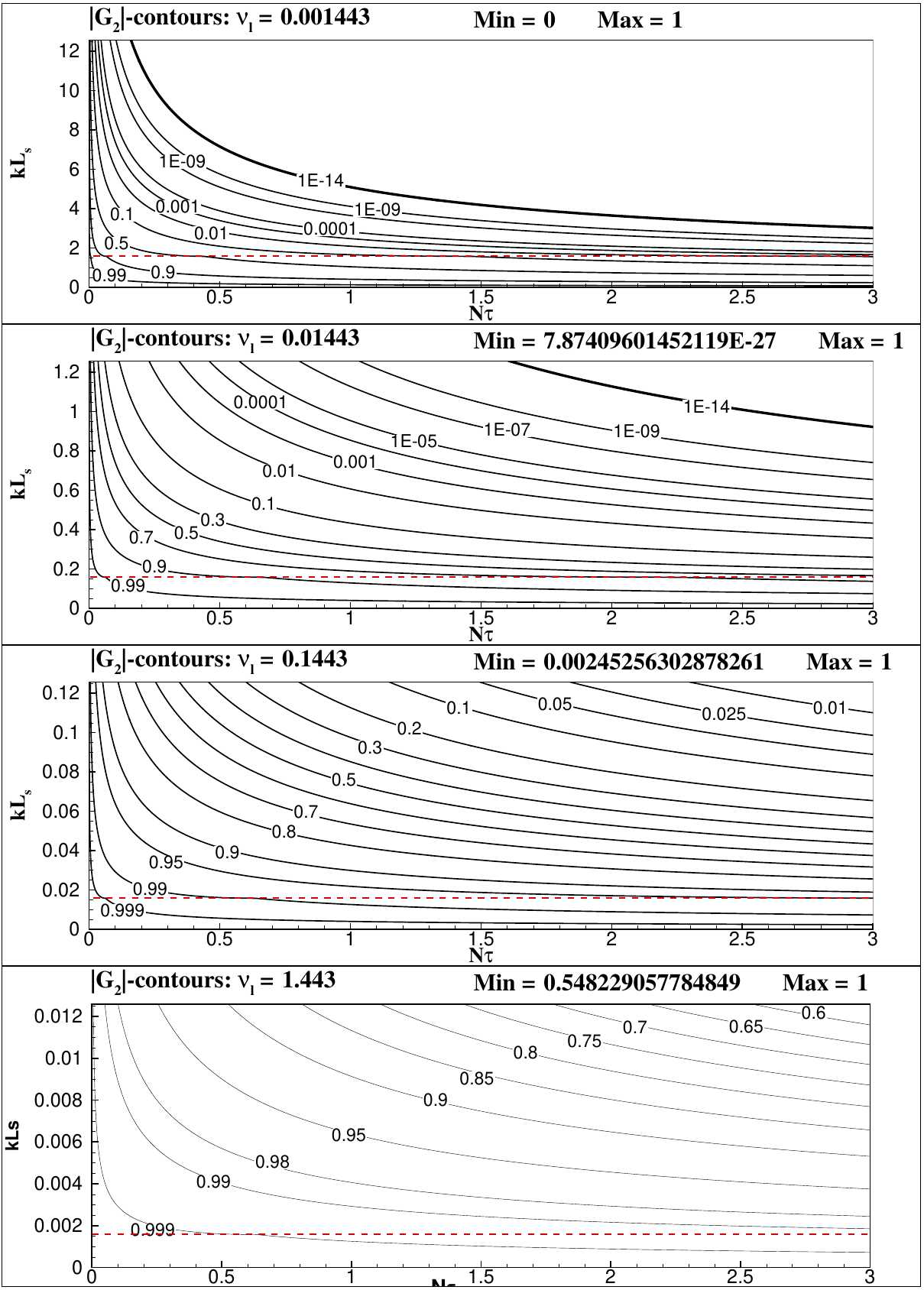}
\caption{The frames show the magnitude of the second mode of the amplification factor for the indicated generalized kinematic viscosity cases in the $(N_\tau, kL_s)$-plane, as given by equation \eqref{Eq:G12_dim}. The length-scale and time-scale have been chosen as in figure \ref{modG1}. The boundary between the regions depicting the diffusion and wave solutions is given by the non-dimensional cut-off wavenumber ($k_c L_s$) shown by a dotted line (red).}
\label{modG2}
\end{figure*}

For the case considered in \cite{Acoustic_POF21} with measurements reported for the attenuated acoustic signal, and the case shown in figure \ref{Figp14}, one notes $c = 343.11 {\rm m/s}$ for air at an ambient temperature of $20^o$C and $\nu_l = 0.1443 {\rm m^2/s}$, which fixes the cut-off wavenumber as $k_c= 4755.568 {\rm m}^{-1}$. The maximum wavenumber ($k_{max}$) is chosen as four times the value of $k_c$ (which can be chosen arbitrarily, in the absence of any physical data). This, in turn helps one to choose the smallest resolved length-scale in this analysis to be given by $L_s = 3.30485 \times 10^{-4}$m. For the fixed diffusion number $D_n = 0.14$, this corresponds to the chosen time-scale to be given by, $\tau_s = 1.0596 \times 10^{-7} {\rm s}$. 

In the following figures, this same value of $L_s = 3.30485 \times 10^{-4}$m is used for all the chosen values of $\nu_l$. In the dimensional form, the amplification factors are given by,

\begin{equation} \label{Eq:G12_dim}
G_{1,2} (k,\tau_s) = e^{-\nu_l k^2 \tau_s/2} e^{\mp i kL_s N_{\tau} \sqrt{1- (k/k_c)^2}}
\end{equation}

The phase speeds are given in the dimensional form as, 
\begin{equation} \label{Eq:c12_dim}
    c_{ph 1,2} = \pm c \sqrt{1- (k/k_c)^2} 
\end{equation}

The corresponding group velocity components are obtained from the real part of the following, 

\begin{equation} \label{Eq:v_g12_dim}
    v_{g 1,2} = \pm c\sqrt{1- (k/k_c)^2} \mp \frac{\left(k \nu_l \right)^2}{4c\sqrt{1- (k/k_c)^2}} - i\nu_l k
\end{equation}

\begin{figure*}
\centering
\includegraphics[width=0.9\textwidth]{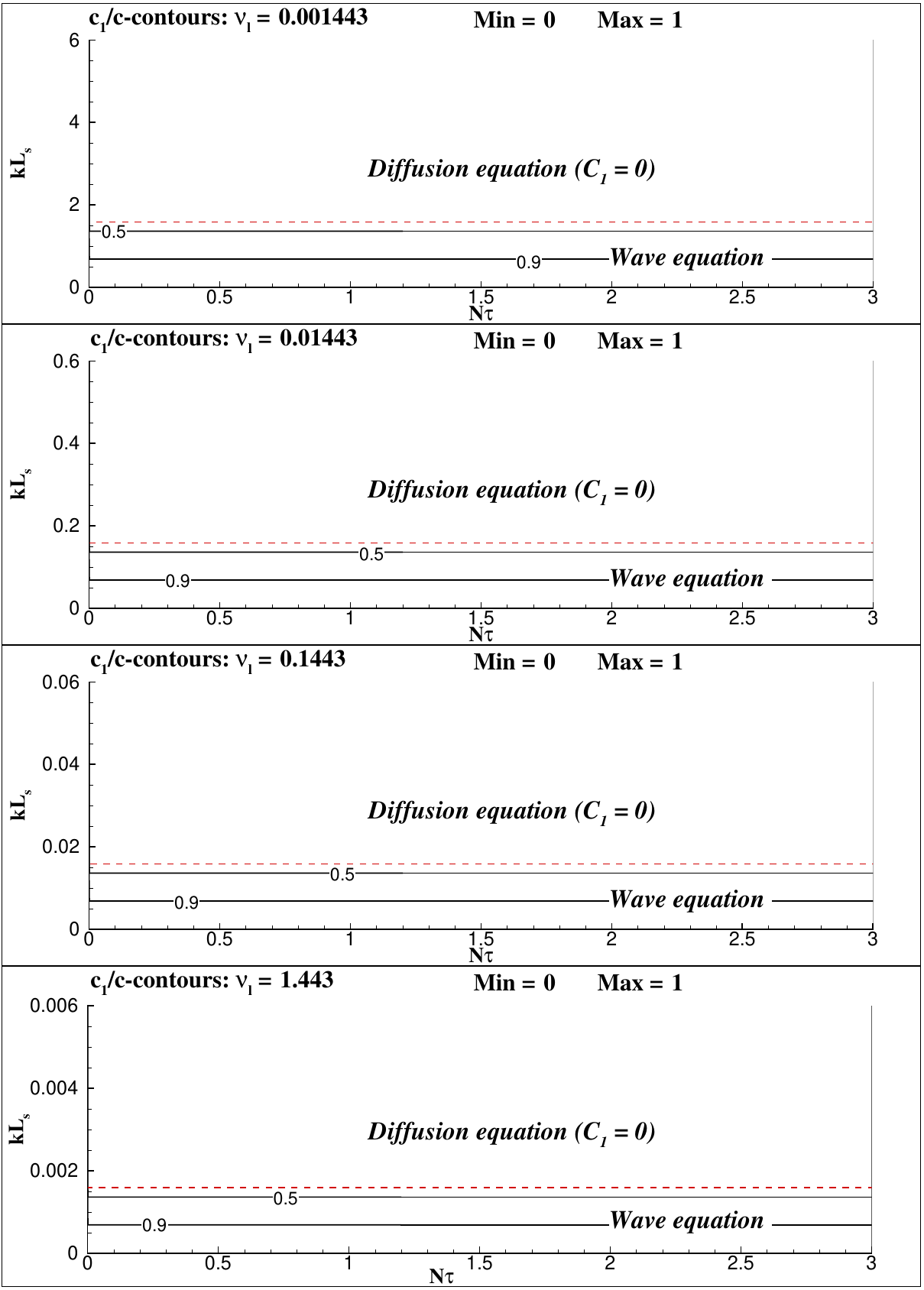}
\caption{The frames show the phase speed of the first mode for the indicated generalized kinematic viscosity cases in the $(N_\tau, kL_s)$-plane, as given by equation \eqref{Eq:c12_dim}. The length- and time-scales have been chosen as in figure \ref{modG1}. The boundary between the regions depicting the diffusion and wave solutions is given by the non-dimensional cut-off wavenumber ($k_c L_s$) shown by a dotted line (red). The phase speed is defined only in the wave solution range.}
\label{c1_dim}
\end{figure*}

\begin{figure*}
\centering
\includegraphics[width=0.9\textwidth]{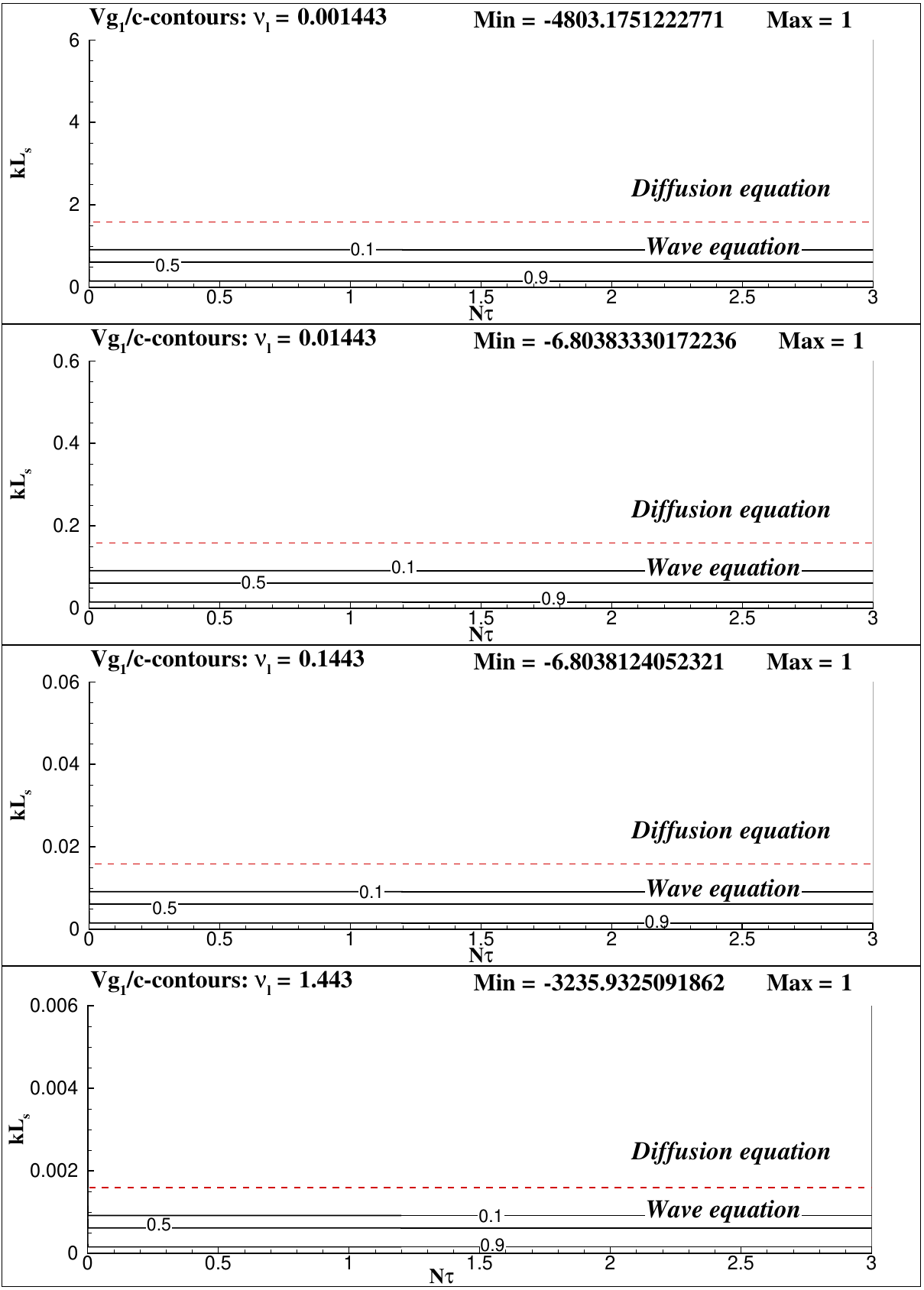}
\caption{The frames show the phase speed of the first mode for the indicated generalized kinematic viscosity cases in the $(N_\tau, kL_s)$-plane, as given by equation~\eqref{Eq:c12_dim}. The length- and time-scales have been chosen as in figure~\ref{modG1}. The boundary between the regions depicting the diffusion and wave solutions is given by the non-dimensional cut-off wavenumber ($k_c L_s$) shown by a dotted line (red). The phase speed is defined only in the wave solution range.}
\label{vg1_dim}
\end{figure*}

It is readily apparent that the phase speed and the physically relevant group velocity (given strictly by the real part) are not functions of $N_\tau$. It has been noted before as in \cite{Acoustic_POF18} that the imaginary part of the group velocity gives rise to anti-diffusion, and thus will have no relevance for physical systems which do not admit anti-diffusion. The expressions given in equations \eqref{Eq:G12_dim}, \eqref{Eq:c12_dim} and \eqref{Eq:v_g12_dim} helps one to plot the solution properties in figures \ref{modG1}, \ref{modG2}, \ref{c1_dim} and \ref{vg1_dim} with $\nu_l$ held as constant.

In figure~\ref{modG1}, the modulus of the first amplification factor, $|G_1|$ is shown for the four generalized kinematic viscosity ($\nu_l$) cases from 0.001443 to 1.443, each increasing by a factor of ten. It is to be emphasized that one is considering the different magnitudes of the coefficient of viscosity, without even bringing into question the utility of Stokes' hypothesis. In all the frames, the line corresponding to $k_cL_s$ are shown by a dotted (red) line. It is to be noted from equation~\eqref{Eq:G12_dim} that for $k > k_c$, the exponent in the second factor becomes purely real, thereby augmenting the first attenuating factor and the amplification factor shows the visible discontinuous jump across the $k_c L_s$-line. For the top frame in figure~\ref{modG1}, this occurs for $kL_s = 1.5857$, and in the subsequent frames below, this value is reduced by a factor of ten. Thus, the $k_c L_s$-line demarcates the mathematical characteristics, where the attenuated wavy solution given by the hyperbolic PDE transforms to the diffusive solution given by the parabolic PDE. In figure \ref{modG2}, the amplitude of the second amplification factor is shown for the same generalized kinematic viscosity cases, which also displays the discontinuity across the $k_c L_s$-line. Above this line, a very interesting different behavior is noted for the two modes, due to the fact that the second exponent in equation~\eqref{Eq:G12_dim}, apart from becoming real, also becomes of opposite sign. Even though the signal is created in a quiescent ambience, the property of the solution is anisotropic with respect to the spatial dimension. It has its root in the viscous term with mixed derivatives of order three. However, for $k < k_c$, the wavy solution is perfectly symmetric. 

In figure \ref{c1_dim}, the phase speed of the first mode is noted as non-trivial wavy solution for $k < k_c$, while for $k > k_c$, the governing equation becomes diffusive in nature with no variation in phase with time, as the imaginary part of the amplification factor is zero. As the phase speed for the second mode is identical in magnitude,but with opposite in sign, this is not shown. In figure \ref{vg1_dim}, the corresponding group velocity is shown for the first mode, with the regions for different $\nu_l$ demarcated between the wavy solution and diffusive solution. The second mode also display similar features with identical magnitude, but having opposite sign for the wavy solution. As $k$ approaches the cut-off wavenumber, for a fixed frequency (or $N_\tau$) the group velocity ceases to exist and becomes undefined.
\subsection{Multi-modal behavior of the governing equation}

If the imaginary part of the physical amplification factor is absent in equation~\eqref{Eq:G1}, then there will be no phase shift in the time interval of $\tau_s$. Such a situation can arise for $\beta_1 = m \pi$, for all integral values of $m$ including zero, i.e. 

\begin{equation} \label{Eq:beta1}
kcf \tau_s = m\pi {\hspace{5mm}} {\rm for}~m = 1,2, .... \infty
\end{equation}

For the general case, terming these wavenumbers as $k_m$, one can rewrite the above condition given by,

\begin{equation} \label{Eq:beta2}
k_m \sqrt{1 - (k_m / k_c)^2} = \frac{m\pi}{c\tau_s} 
\end{equation}

\noindent where the cut-off wavenumber is defined by, $k_c = 2c/ \nu_l$, above which $\omega_{1,2}$ becomes strictly imaginary, and then $G_{1,2} (k, \tau_s)$ will be strictly real, as is noted for parabolic PDEs. For $k_m < k_c$, the circular frequency and the physical amplification factor will be complex, and the spatio-temporal dynamics will display an attenuated wave nature, i.e. the governing equation is given by a hyperbolic PDE, with $G_{1,2}$ as complex conjugates.

With the help of the length-scale, $L_s$, and the non-dimensional time-scale, $N_\tau = \frac{c\tau_s}{L_s}$, the condition given in equation~\eqref{Eq:beta2} can be written alternately as,

\begin{equation} \label{Eq:locus}
N_\tau k_m L_s \sqrt{1 - (k_m / k_c)^2} = {m\pi} 
\end{equation}

One can plot the loci given by equation \eqref{Eq:locus} for different values of $m$ in the $(N_\tau, kL_s)$-plane.

\begin{figure*}
\centering
\includegraphics[width=0.9\textwidth]{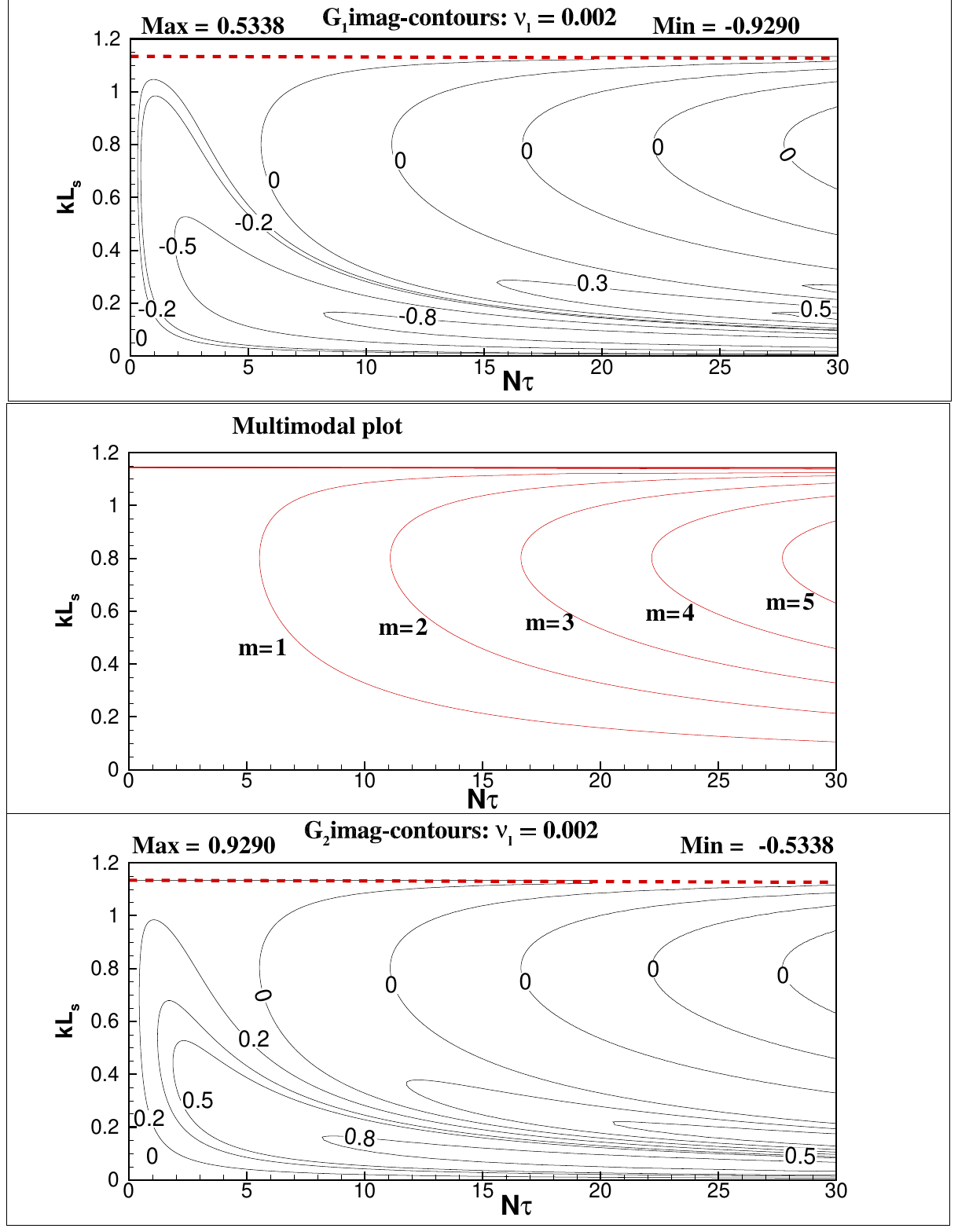}
\caption{The imaginary contours of $G_1$ (top) and $G_2$ (bottom) in extended parameter range of $N_\tau$ showing multi-modal diffusive nature of the perturbation equation. These modes are obtained using equation \eqref{Eq:locus} as shown in the middle frame. For the chosen range of $N_\tau$, one can only notice first five modes identified by $m= constant$ in the figure.}
\label{multi_modalG12}
\end{figure*}

Along these loci, the imaginary part of the amplification factors can be simplified as given next,

\begin{equation} \label{Eq:G12_multi}
G_{1} (k,\tau_s) = e^{-\nu_l k^2 \tau_s/2} e^{-im\pi} \;\;\;\; {\rm and} \;\;\;\; G_2 (k,\tau_s) = e^{-\nu_l k^2 \tau_s/2} e^{im\pi}
\end{equation}

This clearly shows the amplification factors to be strictly real, where the solution will be diffusive, as it has been shown in figure \ref{multi_modalG12} for different integral values of $m$. The locus of points in the $(N_\tau, kL_s)$-plane for the subcritical wavenumbers are shown in the middle frame of this figure, that follows equation \eqref{Eq:locus}. In figure \ref{multi_modalG12}, on the top and the bottom frame show the imaginary part of the amplification factors. It can be noted that one observes multi-modal diffusive nature as described in equation \eqref{Eq:G12_multi}. It implies that solution will be strictly diffusive for multiple such modes for different integer values of $m$. However, the phase speed will only vanish for $m = 0$. 
Along the loci of figure \ref{multi_modalG12} shown in the middle frame the phase shift corresponding to $\beta= \pm m\pi$ will give rise to the phase speed given by, $\beta =k c_{1,2} \tau_s$, i.e. $c_{1,2} = \pm (m\pi)/(k\tau_s)$. 

\begin{figure*}
\centering
\includegraphics[width=0.9\textwidth]{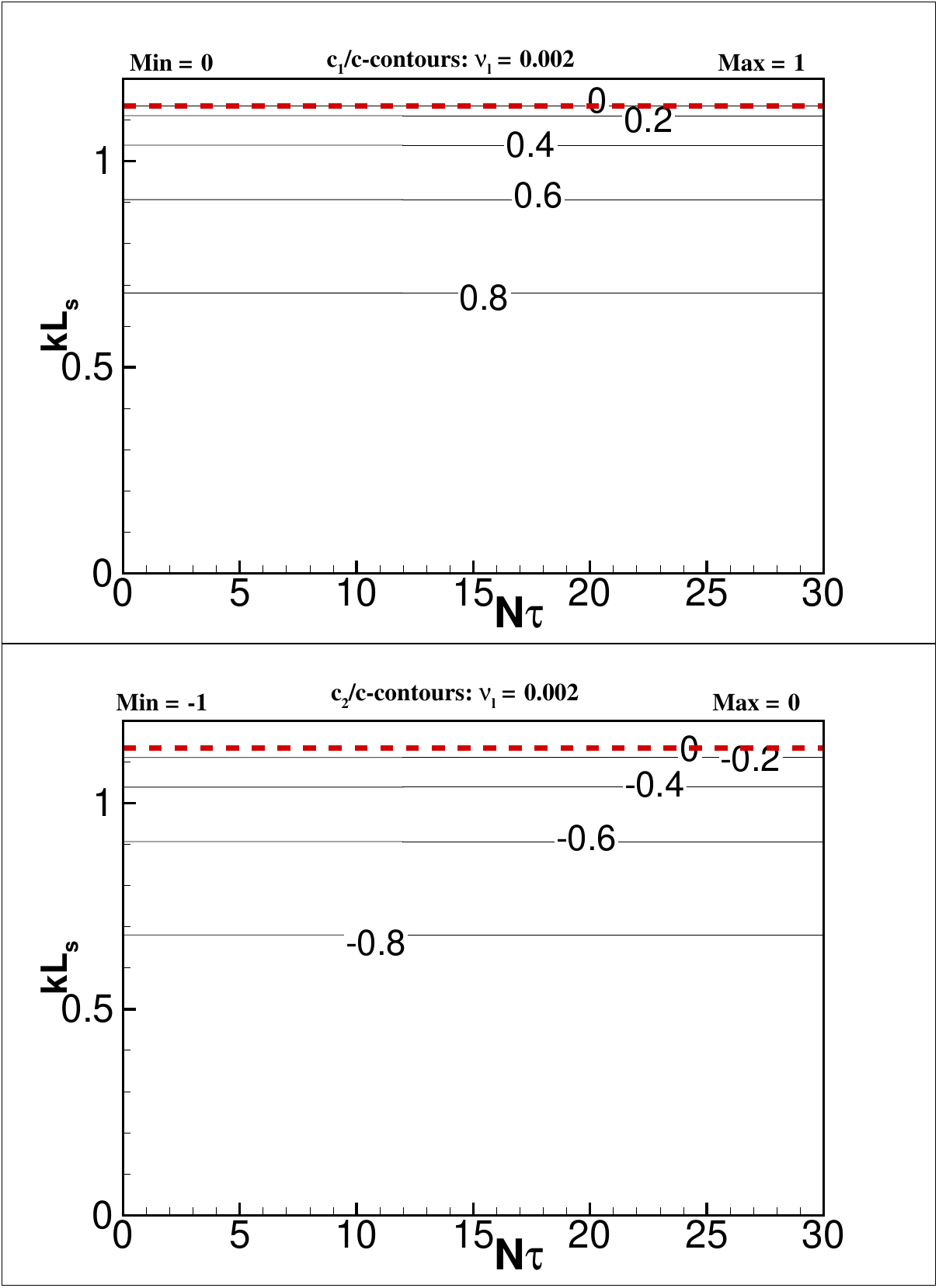}
\caption{The phase speed of both the modes are shown in the extended parameter range of $N_\tau$ up to 30. The phase speed remains independent of $N_\tau$, and the multi-modal diffusive nature of the perturbation equation noted in figure \ref{multi_modalG12} will not be visible here. The phase speed is given by $\pm (m\pi)/(k\tau_s)$.}
\label{multi-modalcph}
\end{figure*}

\begin{figure*}
\centering
\includegraphics[width=0.9\textwidth]{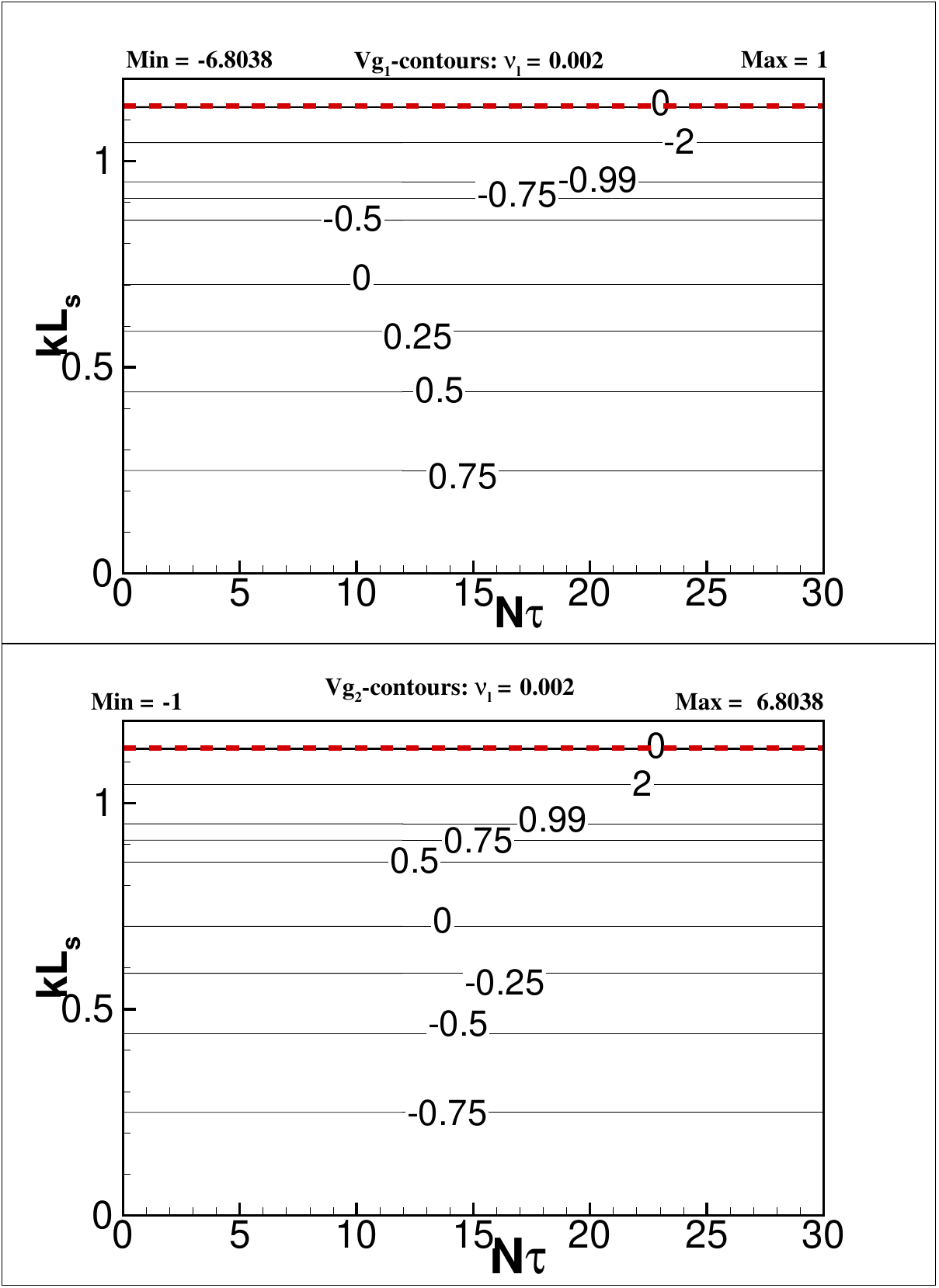}
\caption{The group velocity of both the modes are shown in the extended parameter range of $N_\tau$ up to 30. The group velocity is also independent of $N_\tau$. The multi-modal diffusive nature of the perturbation equation noted in figure \ref{multi_modalG12} is noted here for two such modes in the plotted contours.}
\label{multi-modalvg}
\end{figure*}

As the phase speed is depicted over the extended time-scale range in figure \ref{multi-modalcph}, the corresponding group velocity plots are shown in figure \ref{multi-modalvg}. Here also, one notices vanishing values of the group velocity corresponding to two values of $kL_s$ in the figure. 

\section{Summary and Conclusions}
The present work uses of global spectral analysis in the theoretical framework of acoustic disturbance propagation in a quiescent ambience. It is obtained by splitting the field into a mean and its perturbation components and thereafter linearizing the compressible Navier-Stokes equation. The analysis helps one to distinguish between the wave-like propagation of the disturbance and associated diffusion equation depending upon the length scale, for a given time scale. The existence of a cutoff wavenumber, $k_c$, that demarcates the wave equation and the diffusion equation is a novel feature of the reported research here. This exact quantification for a one-dimensional planar signal, has similarity with the Kolmogorov's length scale which also postulates the conversion of kinetic energy (of the wave equation) to heat energy diffusing at very small scales. The dispersive nature of the governing equation raises the important question about the speed of sound for the dispersive phenomenon. The role of the generalized kinematic viscosity, $\nu_l$ is responsible for the observed dispersion and showing the same governing equation to represent wave motion for some lower wavenumber to diffusion at a higher wavenumber and beyond. We also demonstrate further the existence of gaps in the plane constituted by a non-dimensional time scale ($N_{\tau}$), and nondimensional wavenumber ($kL_s$) when the governing equation is viewed in the enlarged domain. A multi-modal behavior is noted for such large ranges of the independent variable when one considers fluids with very small values of $\nu_l$. The current research suggests that a careful measurement of the numerical properties of the acoustic wave propagation is essential, especially for ascertaining the roles of phase speed and group velocities of the signal.

\bibliographystyle{jfm}
\bibliography{acoustics_edited}


\end{document}